\documentclass[10pt,journal,compsoc]{IEEEtran}
\ifCLASSOPTIONcompsoc
  \usepackage[nocompress]{cite}
\else
  \usepackage{cite}
\fi
\ifCLASSINFOpdf
\else
\fi

\usepackage{amsmath}
\usepackage{graphicx}

\usepackage{amsmath}
\usepackage{amssymb}
\usepackage{diagbox}
\usepackage{slashbox}
\usepackage{soul}
\usepackage{subfigure}
\usepackage{multirow,tabularx} 
\usepackage{xcolor}

\newcommand{\D}{\mathcal{D}}

\begin{document}
%
\title{A Survey of Data Pricing for Data Marketplaces}
%
%
%
%

\author{Mengxiao~Zhang,
        Fernando~Beltr\'an,
        and~Jiamou~Liu\textsuperscript{*}
\IEEEcompsocitemizethanks{\IEEEcompsocthanksitem Mengxiao Zhang is with School of Computer Science and Engineering, University of Electronic Science and Technology of China, Chengdu, Sichuan, CN. 
E-mail: mengxiao.zhang@uestc.edu.cn
\IEEEcompsocthanksitem Fernando Beltr\'an is with Business School, The University of Auckland, Auckland, NZ. Email: f.beltran@auckland.ac.nz.
\IEEEcompsocthanksitem Jiamou Liu is with School of Computer Science, The University of Auckland, Auckland, NZ. Email: jiamou.liu@auckland.ac.nz.
}
\thanks{This paper is partially supported by National Natural Science Foundation of China No. 62172077.}}

%
%

\markboth{Journal of \LaTeX\ Class Files,~Vol.~14, No.~8, August~2015}%
{Shell \MakeLowercase{\textit{et al.}}: Bare Demo of IEEEtran.cls for Computer Society Journals}
%



\IEEEtitleabstractindextext{%
\begin{abstract}
A data marketplace is an online venue that brings data owners, data brokers, and data consumers together and facilitates commoditisation of data amongst them. Data pricing, as a key function of a data marketplace, demands quantifying the monetary value of data. A considerable number of studies on data pricing can be found in literature. This paper attempts to comprehensively review the state-of-the-art on existing data pricing studies to provide a general understanding of this emerging research area. Our key contribution lies in a new taxonomy of data pricing studies that unifies different attributes determining data prices. The basis of our framework categorises these studies by the kind of  market structure, be it sell-side, buy-side, or two-sided. Then in a sell-side market, the studies are further divided by query type, which defines the way a data consumer accesses data, while in a buy-side market, the studies are divided according to privacy notion, which defines the way to quantify privacy of data owners. In a two-sided market, both privacy notion and query type are used as criteria. We systematically examine the studies falling into each category in our taxonomy. Lastly, we discuss gaps within the existing research and define future research directions.
\end{abstract}

}

\maketitle

\IEEEdisplaynontitleabstractindextext

%
\IEEEpeerreviewmaketitle

\IEEEraisesectionheading{\section{Introduction}\label{sec:introduction}}

%
%
%
%

\IEEEPARstart{``T}{he} world's most valuable resource is no longer oil, but data'' as proclaimed by The Economist in 2017\footnote{Parkins, David. "The world’s most valuable resource is no longer oil, but data." The economist 6 (2017).}. 
The advancements of information technologies such as Web $2.0$, cloud computing, and Internet of Things have fuelled an unprecedented availability of data. 
The increasing adoption of data-driven decision making by individuals, businesses, and governments has unlocked the vast economic value of data. Indeed, some of the world's largest and most successful companies such as Alibaba and Alphabet have data at their core business. In the healthcare sector of the US, for example, data-driven technologies generate additional \$300 billion in value each year, 2/3 of it comes from reduced medical expenses \cite{manyika2011big}. A comparable level of value generated by data is observed in other areas such as retailing, government administration, and finance \cite{fisher2018using,maciejewski2017more,yin2015big}.  
%
%
A major portion of the input to data-intensive technologies comes from private data, i.e., information that is usually expected to be withheld by individuals and out of the public eye. 
On one hand, privacy is regarded as a basic personal right and the lack thereof contributes directly towards one's loss of security and wellbeing. On the other hand, some level of privacy must be compromised due to data mining for the masses. This {\em inherent conflict between upholding individual rights and advancing collective utility} presents the central challenge in today's data-intensive technologies. 

In the last three to five years, the emergence of {\em data marketplaces} presents an opportunity to resolve this conflict. 
Through the monetisation of data, a data marketplace aims to strike a balance between individuals' privacy loss and the utilisation of private data. 
For instance, the Australia-based company Data Republic\footnote{https://www.datarepublic.com/} provides de-identified bank transactional data to a fast food chain. Combining the external data with the internal data, the fast food chain developed better promotional strategies than their competitors\footnote{https://www.datarepublic.com/resources/team-blog/project-highlights-across-our-data-ecosystem}. In another example,  the UK-based Citizenme\footnote{https://www.citizenme.com} collects data for mobile phone brand Wileyfox to improve their business strategy\footnote{https://www.citizenme.com/case-studies/measuring-brand-nps/}. Other real-world examples of data marketplace businesses include AggData\footnote{https://www.aggdata.com}, 
Data \& Sons\footnote{https://www.dataandsons.com}, Snowflake\footnote{https://www.snowflake.com/}, Datum\footnote{https://datum.org/}, digi.me\footnote{https://digi.me} and Lotame\footnote{https://www.lotame.com/}. 
On one side of the market, {\em data owners} such as individuals or organisations who have private data hold their personal information and are open to exchanging it for an amount of compensation, which should reflect their loss of privacy as a result of the exchange.  On the other side of the market, {\em data consumers} such as advertisers, retailers, manufacturers, and service providers seek private data for analytical tasks and are willing to pay for the data for their generated economic value. A data marketplace is a type of online venue that brings data owners and data consumers together and facilitates the buying and selling of data between them. 
In many situations,  a {\em data broker} would act as an intermediary between data owners and data consumers, collecting, integrating, storing, redistributing data, and generating profits from such operations.
The data broker could either be an integral component of the platform or an external third-party. 
{\em Data pricing} aims to quantify the monetary value of data and is a key function of a data marketplace. 
For example, AggData sells location data at fixed prices, bundles of datasets at a discount, unrestricted access to all datasets for a subscription fee, and custom data by negotiation. On the other hand, Datacoup offers individuals a monthly fee if they allow other users to access the information of their online accounts, such as Facebook, LinkedIn, and Google+. These various data pricing strategies require systematic studies into the demand and supply of data and how the value of data is materialised under different settings. 

The market structure plays a key role in determining the value of data. 
Data marketplaces differ by their market structures. There are mainly three types of data marketplaces,  sell-side market, buy-side market, and two-sided market \cite{stahl2016classification}. 
A {\em sell-side market} integrates data from multiple sources and sells data to data consumers. 
For example, AggData offers a sell-side market for location data. 
On the other hand, a {\em buy-side market} enables individuals and organisations to monetise their internal data, allowing data brokers to procure data from data owners. 
A {\em two-sided market} combines the two sides by allowing data owners to provide their data to data consumers. E.g., Data \& Sons offers a two-sided data market.
On the sell side, the monetary value of data is decided by how much additional value the consumers expect data to generate. 
On the buy side, the value of data captures how much privacy is given up by the data owner as a result of the data trade. 
When two sides are connected by a data broker, the price of data should be indicative of the considerations of both the data consumers and data owners.
Naturally, data owners with lower privacy requirements should be better compensated and data consumers with higher information requirement should be charged higher.  

To capture privacy requirements of data owners, {\em differential privacy}, proposed by Dwork in \cite{Dwork2006Differential}, offers a sweeping theoretical framework. 
The notion rigorously defines the privacy using the probability one can infer an individual data record from aggregated query outputs.
This definition and its variants have attracted intense investigation and industry uptake \cite{dwork2014algorithmic,cormode2018privacy}. 
%
On the other hand, differential privacy is by far not the only way to capture privacy. For example, Gkatzelis et al. \cite{gkatzelis2015pricing} quantify privacy by the number of data consumers who access the data. Parra-Arnau \cite{parra2018optimized} quantify privacy by the rate of disclosing the ground truth record. These (non-DP) ad hoc definitions of privacy usually focus on specific application scenarios or data types. Nevertheless, they form a non-trivial literature in analysing data values for buy-side markets.  

To capture information requirements of data consumers,  we broadly categorise queries that data consumers demand into three types. Firstly, in many data marketplaces such as AggData, the dataset is directly accessed by the data consumer after the data trade. We refer to this type of data request a {\em null query}. 
Then, a {\em one-off query} returns a single and fixed aggregate statistic, such as sum, average, minimum, and maximum of the data. For example, BookYourData\footnote{https://www.bookyourdata.com/} and DirectMail\footnote{https://www.directmail.com/} allow data consumers to use a query to target the records with attribute values of interest. 
Last, in a {\em general query}, a data consumer may issue multiple queries and the query answer is an aggregate statistic or a view consisting of multiple rows and columns. 

A considerable number of studies on data pricing methods exist in the literature. The research on data pricing is initiated by Jaisingh et al. \cite{jaisingh2008privacy} from the perspective of operations research. They treat private data as a special case of information products and the privacy level of data is a determinant of its price. This idea is adopted by Riederer et al. \cite{riederer2011sale}, who studies unlimited supply auction for the selling of datasets. 
%
In the same year, Ghosh and Roth \cite{ghosh2011selling} reconsider the privacy issue under one-off queries.
Different from \cite{jaisingh2008privacy}, they use differential privacy to quantify the privacy loss of data owners and links it with the value of data. 
The assumptions made in \cite{ghosh2011selling} are relaxed in a series of subsequent studies \cite{roth2012conducting,fleischer2012approximately,dandekar2012privacy,ghosh2014buying}. 
In a different vein, \cite{balazinska2011data,koutris2012query} discuss the arbitrage problem when selling data under general queries and provide an arbitrage-free pricing scheme to address this issue. The arbitrage issue in pricing general queries has attracted much academic attention since then. The solution in \cite{koutris2012query} is restricted to a specific query type and solutions for more general queries are proposed in subsequent studies \cite{Lin2014On,Deep2016The,Tang2013The}. 
Li et al. \cite{Li2014A} extend \cite{koutris2012query} by considering the privacy and arbitrage issues at the same time. 

Muschalle et al. \cite{muschalle2012pricing}, Schomm et al. \cite{schomm2013marketplaces} and Stahlet al. \cite{stahl2014data} were among the first surveys on data marketplaces and data pricing. These surveys empirically investigate the characteristics of data marketplaces, their participants and pricing models used in practice. They, however, do not provide a comprehensive overview of the topic as they fail to discuss key issues such as information asymmetry in data trading. 
Liang et al. \cite{liang2018survey} conduct a survey on the studies related to big data market. The focus of \cite{liang2018survey} is on the life cycle of big data rather than pricing itself.  
Wagner et al. \cite{wagner2018putting} systematically review the empirical studies of the factors that affect the perceived valuation of personal information. 
It provides insights for determining the value of data, but does not mention the work on data pricing methods. 
Moreover, Fricker and Maksimov \cite{fricker2017pricing} provide a review of the pricing models and the pricing items in $18$ research papers. They use a classification that can be generally applied to the trading of arbitrary products while omitting the unique characteristics, such as the need for privacy preservation, of data trading. In particular, this classification does not reflect the valuation of data. 
The most recent literature review is from  Pei in \cite{pei2020survey} which summarises desirable properties of data pricing strategies that were considered in existing studies. The dimensions considered include, e.g., truthfulness, revenue maximisation, fairness, arbitrage-freeness, privacy preservation, and computational efficiency. However, none of the survey above synthesises an integrative and unified framework by linking the various factors of data pricing.

This paper attempts to comprehensively review the state-of-the-art on existing data pricing studies to provide a general understanding of this emerging research area. Our key contribution lies in a new taxonomy of data pricing studies that unifies different attributes determining data prices.
The basis of our framework categorises these studies by the kind of {\em market structure}, be it sell-side, buy-side, or two-sided, where data trade is assumed to take place. 
(1) In a {\em sell-side market}, the value of data lies in the value of information derived from data for data consumers. Information are provided to a consumer through queries, and query type shapes how the value of information is measured. This suggests the use of {\em query type} as the key criterion to distinguish data pricing studies.
(2) In a {\em buy-side market}, data represents data owners' privacy. Under different privacy notions, the privacy loss of data owners are measured in different ways. Therefore {\em privacy notion} is selected as the criterion. 
(3) In a {\em two-sided market}, data hold value in the two forms at the same time, and both query type and privacy notion determine the ways of measuring data value. Thus both of these two criteria are considered. 
%
We systematically examine the studies that fall into each category in our taxonomy framework. 
We then review the work discussing economics of privacy in the field of economics and management. 
Lastly, we discuss gaps within the existing research and define future directions. Firstly, it is the potential to apply machine learning and reinforcement learning to assist solving complex data pricing problems. Further, most of the existing studies are confined to relational tables and queries on them. Data in other forms, such as natural language, requires novel pricing schemes.

This paper proceeds as follows: Section \ref{sec:data marketplaces} describes data marketplaces. Section \ref{sec:Overview} introduces the fundamentals of economic and pricing models. 
Section \ref{sec:taxonomy} discusses the taxonomy of studies on data pricing according to market structure, query type and privacy notion, which is followed by the detailed review of the data pricing studies in each category in Sections \ref{sec:sell}, \ref{sec:buy}, \ref{sec:two}. Section~\ref{sec:eco} review the studies on private data value in the field of economics and management. To close, Sections \ref{sec:future} and \ref{sec:conclusion} presents future opportunities and a conclusion.

\section{Data marketplaces}
\label{sec:data marketplaces}

Although the idea of trading of digital products, i.e., information, was considered as early as 1998 in the seminal work of Armstrong and Durfee \cite{armstrong1998mixing}, the term ``data marketplace'' was coined by Schomm, Stahl, and Vossen in 2013 \cite{schomm2013marketplaces}. There, the authors defined a data marketplace as a platform where users purchase different licenses for accessing various datasets. Subsequently, Muschalle, Stahl, L\"{o}ser, and Vossen \cite{muschalle2012pricing} argued that data marketplaces not only help to connect demand with supply, but also play a key role in collecting and storing data, as well as unifying data format. More recently, Stahl and Vossen \cite{stahl2016classification} extended the function of data marketplaces from trading datasets to trading any data-related services (e.g., those that involve data mining algorithms). 

\begin{figure*}[h]
  \centering
  \includegraphics[width=0.8\linewidth]{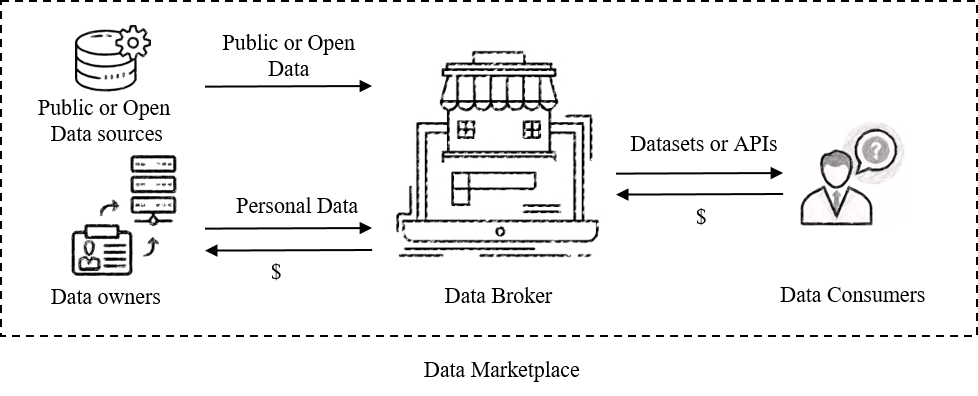}
  \caption[Data marketplace business model]{Data marketplace business model (adopted from \cite{muschalle2012pricing})}
  \label{Fig:BusinessModel}
\end{figure*}

Data marketplaces can be extracted as a business model as shown in Figure~\ref{Fig:BusinessModel} \cite{muschalle2012pricing}. There are mainly three types of participants, data owner and data consumer, data broker.
{\em Data owners} are those who own data and are willing to monetise their data, such as individuals who have social, financial, location, health data about themselves, and companies who collects data about their users, including their demographics, preferences and behaviours. {\em Data consumers} are those who seek for external data to improve their decisions, product design, services and customer management, such as advertisers, software developers, retailers, manufacturers, telecommunication service providers. A {\em data broker} is an intermediary, procuring data from data owners, integrating open data, and selling data to data consumers, and it makes profit from data trade. The data companies listed above work as data brokers. 

\subsection{Structure of data marketplaces}

\begin{figure*}[h!]
\centering
\subfigure[Buy-side market]{\label{subfig:a} \includegraphics[width=0.29\linewidth]{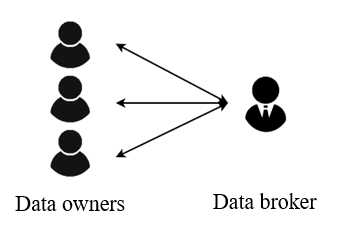}} 
\subfigure[Sell-side market]{\label{subfig:b} \includegraphics[width=0.30\linewidth]{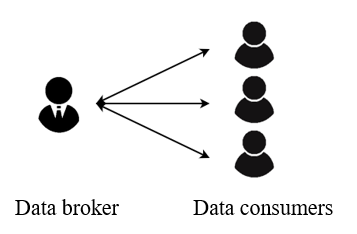}} 
\subfigure[Centralised two-sided market]{\label{subfig:c} \includegraphics[width=0.48\linewidth]{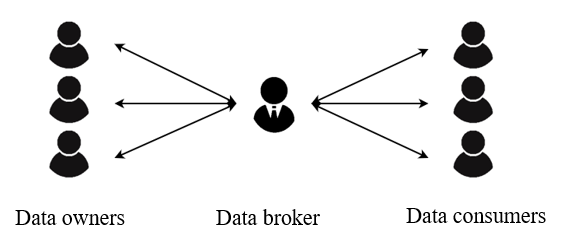}} 
\subfigure[Decentralised two-sided market]{\label{subfig:d} \includegraphics[width=0.4\linewidth]{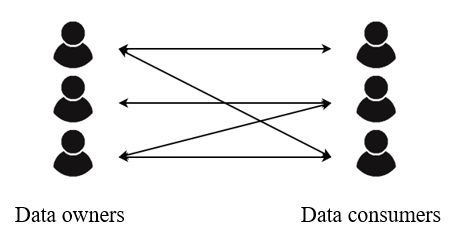}}
\caption{Four data market structures}
\label{fig:structure}
\end{figure*}

While the businesses operated by a data broker include data procurement and data selling, their core business varies across different data marketplaces. Depending on the number of bilateral transactions occurring at a data exchange, the market structure, according to \cite{stahl2016classification}, is known as:

\begin{itemize}
    \item {\em One-sided market}: a model of market interaction in which a data marketplace deals with either a buy-side or a sell-side, but not both. 
    \begin{itemize}
        \item {\em Buy-side market}: In a buy-side market, as shown in Figure~\ref{subfig:a}, a data broker collects data from multiple data owners and compensates them accordingly. 
        A buy-side market serves organisations and individuals who want to monetise their data.  
        \item {\em Sell-side market}: Sell-side market deals with the data transactions between a data broker and data consumers, as shown in Figure~\ref{subfig:b}. For organisations and individuals who seek external data sources for analytical tasks \cite{Forrester2017}, a sell-side market serves to meet their data demand.
    \end{itemize}
    \item {\em Two-sided market}: a model of market interaction in which a so-called platform attends to both, the buy-side and the sell-side. A data marketplace can be regarded as a two-sided market, where an intermediary - the data broker - provides a data trading platform for data owners and data consumers and profits from data transactions.
    \begin{itemize}
        \item {\em Centralised two-sided market}: in a centralised two-sided market, all data owners and data consumers trade data through a data broker, as shown in Figure~\ref{subfig:c}. The data broker needs to decide both the procurement prices for data owners and the selling prices for data consumers. The profit of the data broker is the difference between the revenue generated from data selling and the costs incurred by data collection. For instance, Data \& Sons, as a data broker, monetises data owners' personal data and sells aggregate information to data consumers.  
Since only aggregate query results are sold to data consumers, the company would preserve the personal privacy of data owners.  
        \item {\em Decentralised two-sided market}: in a decentralised data market, as shown in Figure~\ref{subfig:d}, a data broker provides a trading platform for data owners and data consumers, but do not involved in the data trade while data consumers and data owners make transactions directly if they are members of the data trading platform. Datum, CitizenMe and DataWallet\footnote{https://datawallet.com/} are examples of data marketplaces with such a market structure. 
    \end{itemize}
\end{itemize}

\subsection{What does a data marketplace trade on?}


{\bf Sell side.}
There are two types of a commodities on the sell side: datasets and queries. 
Datasets can be structured, such as relational tables, or unstructured, such as images, audio, graphics and text. 
Most of the existing works assume that a dataset corresponds to one or more relational tables. In a table, every column describes an attribute while every row represents a record, which is associated with an object. For instance, the locational data sold on AggData are in such form that a dataset is a table with multiple columns and rows that describes business locations in a certain industry. Each column corresponds to an attribute, such as address, city, state, zip code, phone number, etc. Each record corresponds to a specific business location. 
In contrast to datasets, queries answer specific questions from data consumers. A query answer can be a view, which is typically made up of combinations of rows and columns of a dataset. For instance, DirectMail\footnote{https://www.directmail.com/} sells mailing lists of customers and businesses. Their data consumers can use filters for attributes such as zip code, age, gender, etc, to select the targeted customers or businesses. Also, query answers may also involve aggregate information and summary measurements such as sum, average, minimum, and maximum of the data or particular subsets of data.

{\bf Buy side.} 
A  data broker collects data from individual data owners on the buy side. A data owner can be a company that has a dataset of its users for sale or an individual who has a record about herself. In either case, the data traded on the buy side are personal data.


\section{Overview and fundamentals of pricing methods in data transactions}
\label{sec:Overview}
This section introduces basic economic and pricing models, which are adopted in data pricing. These  concepts lay the foundation for the analysis of data pricing studies in subsequent sections.

On the sell side, the maximum price that a consumer can offer to get the product, or willingness-to-pay (WTP), is considered when setting prices for data consumers. While on the buy side, the minimum price that a data owner is willing to be compensated for in exchange for sharing her data, or willingness-to-accept (WTA), matters. The knowledge of the data broker also plays a crucial role in data pricing. There are two cases to consider: (1) when the data broker has {\em complete information}, i.e., there is a good knowledge about the preferences of data owners and consumers,  and (2) when the data broker has only {\em asymmetric information}, i.e., the preferences are only known by the data consumers (or the data owners) themselves. 
The pricing approaches in the settings with complete information and asymmetric information are discussed in the following two subsections, respectively.  

\subsection{Pricing methods in contexts of complete information}
\label{sec:com}
A data marketplace may apply several pricing methods such as flat-fee, premium, pay-per-use, two-part-tariff, versioning, and bundling. The methods above assume that the preferences of data consumers or data owners are known by data brokers, 
thus optimal prices can be achieved accordingly. 

    \textbf{Flat-fee}. A flat fee usually means a monthly or yearly subscription fee associated with a usage limit. In the flat-fee pricing strategy, time is the only factor determining how much a data consumer needs to pay.
    For instance, Infochimps charges a monthly subscription fee and allows data consumers to call a certain amount of queries \cite{balazinska2011data}. It is one of the simplest data pricing methods. 
    
    \textbf{Premium}. Premium is a special case of subscription options. It means monthly or yearly subscription fee associated with unlimited usage. In the premium model, there is no restriction on the number of datasets or queries a consumer can access. For example, Aggdata gives consumers who pay a premium fee unrestricted access to its more than $4,500$ datasets. 
    
    \textbf{Pay-per-use (or linear pricing)}. Pay-per-use or linear pricing is a flexible pricing method, which is based on usage volume. The price $p(n)$ is a linear function of the usage amount $n$, i.e., $p(n)=p_u n$
    where $p_u$ is the unit price of products. Compared to flat-fee, a distinct advantage of pay-per-use is its flexibility. It is preferable for non-frequent consumers, who buy data once or just a few times and leave the market for good.
    
    \textbf{Two-part-tariff}. Two-part tariff is a combination of flat fee and pay-per-use \cite{shapiro1998information,shy2008price}. 
    When it is applied in data marketplaces, data consumers pay a fixed fee for a certain amount of datasets or queries firstly. If they have used up the quota and want more data, they can buy data at a unit price, i.e., $p(n)=p_f+p_u n$
    where $p_f$ is the fixed part and $p_u n$ is the variable part. It suits consumers with uncertain demands.
    
    \textbf{Tiered pricing}. Tiered pricing sets $k$ different unit prices for different quantities of an item \cite{kushal2012pricing}. In other words, it has $k$ levels of amounts $t_j$ and prices $p_j$. The pairs of amount levels and prices are $(t_1,p_1),\ldots,(t_k,p_k)$. For each level, $p(m)=p_j m$, where $t_{j-1}  \leq m < t_j$, $j=2,\ldots,k$ with $t_1 \leq t_2 \leq \ldots \leq t_k$ and $p_1 \geq p_2  \geq \ldots \geq p_k$. With an increase in the amount, the unit price decreases, which indicates that there is a quantity discount that promotes sales in some sense. In tiered-pricing methods, both the price $p_j$ and the length of the interval between $t_{j-1}$ and $t_j$ need to be determined.
    
    \textbf{Versioning}. Versioning is providing multiple versions of information products at different prices \cite{shapiro1998versioning}. Versions may vary in functions, convenience, and speed operation, amongst others, and serve different kinds of consumers. For example, in data marketplaces, data can be sold in multiple versions with different qualities or quantities. For instance, Quandl\footnote{https://www.quandl.com}, a financial and economic data company, provides four versions of their datasets. 
    
    \textbf{Bundling (or Packaging)}. Bundling is a special case of versioning. It puts more than one items together and sells them as a whole package. The price of the package is usually lower than the sum of the prices of the items involved. Data marketplaces apply bundling pricing method as well. AggData, as an example, packages $12$ lists of small American cuisine restaurants and sells it at a discount price, $\$ 425$, while the unbundled lists would cost $\$ 688$. 

\subsection{Pricing methods in contexts of asymmetric information}

{\em Information asymmetry} refers to a common issue during a transaction where one party holds information that is not held by the others, leading to inefficient allocation of resources. 
In a data marketplace, on the sell-side, a data consumer holds an expected value of the data to be purchased, yet the information is not known by the data broker. In this case, the data consumer can strategically pay less than the true value to increase utility. 
To reduce economic losses due to the this information asymmetry, the data broker needs to provide some incentives for data consumers  so that they will be willing to reveal their hidden information. Contracts, auctions and incentive mechanisms are methods used to solve the information asymmetry problem. 

The data consumers and data owners are rational agents who act strategically. In other words, given the trading rules, they make predictions about the others’ strategies and choose the optimal action accordingly. To reveal the hidden information of their valuations, the data broker needs to make sure that for data owners, revealing the true valuation is an optimal strategy and participating in the transactions does not disadvantage them. The former condition is known as {\em incentive compatibility} (IC), whereas the latter is known as {\em individual rationality} (IR) \cite{borgers2015introduction}. Further, a data broker often has a budget for data collection. When the total payment for all data owners does not exceed the budget, the {\em budget feasibility} (BF) condition is satisfied. 


    \textbf{Contract.} In contract theory, an agent is doing some works for a {\em principal} and their interests are often in conflict. The agent has more information than the principal. However, the latter can not enforce the former reveal her information. Instead, the principal offers a menu of well-designed contracts to the agent and ask her to select one of them. Each of the contracts shows a possible action the agent may take and the corresponding reward. By observing the results caused by the agent’s actions, the principal can infer some hidden information.
    Insurance contract is a typical example of contract design. An insurance company does not know the health condition of a potential insured, which is hidden information. But by offering a menu of insurance contracts and observing her choice, the company can infer some health information \cite{borgers2015introduction}.
    
    In data transactions, the principal can be a data broker and the agent can be a data owner or a data consumer. Data owners often have different preferences on privacy and data consumers have different preferences on data. The data broker provides a menu of contracts consisting of different data and prices and lets data consumers or data owners choose an offer. 
    
    \textbf{Auction}. Auctions have been used to determine prices in practice since they were ever used. Most common examples of auctions are the English ascending auction where multiple bidders respond to prices cried out by an auctioneer as they increase. Formally, an {\em auction} defines a set of rules used to allocate resources and set prices based on information from agents \cite{mcafee1987auctions}. In an auction, an {\em auctioneer} seeks to sell or purchase one or more items via a well-defined auction. A set of {\em bidders} follow the auction rules and each of them responds by revealing, at prescribed moments in time, a value akin to their willingness to pay for (or purchase) a subset of the items on sale. The auctioneer conducts the auction and announces the final price when conditions for closing the auction are met, which triggers an allocation of the items (or rights to sell) to winning bidders. 
    
    An auction in data marketplaces works as follows. A data broker (auctioneer) declares allocation and pricing rules. data owners or data consumers (bidders) submit their bids, which might be open or sealed. Then the data broker announces who the winner(s) is (are) and how much the winner(s) need(s) to pay or to be paid on the basis of predefined rules and submitted bids.
    
    \textbf{Incentive mechanism}. More generally, auctions fall within a larger collection of techniques studied in {\em mechanism design}, which provides a general framework to resolve information asymmetry. 
    Loosely speaking, mechanism design amounts to an inverse version of the problem of deriving equilibria of a game. Specifically, in a game where players act strategically, given desired outcomes, which is defined by an equilibria of the game, mechanism design seeks the rules of the game that determines the outcomes. An {\em incentive mechanism} defines the action sets of agents and rules by a function mapping from action sets to outcomes. 
    In a mechanism with two properties of IC and IR, the players are incentivised to participate in and act according to their true valuations. 
    
    Two main solution concepts have been adopted in mechanism design research, {\em dominant strategy equilibrium} and {\em Bayesian Nash equilibrium}. In a Bayesian Nash equilibrium, the preferences are hidden information, but the distribution of the preferences in the market is commonly known to all agents, and each agent maximises her expected utility based on her knowledge. On the other hand, a dominant strategy equilibrium is a stronger solution concept where there is no assumption about preference distributions. When a dominant strategy exists, the agent will choose such a strategy that maximises her utility regardless of the other agents’ strategies. Under these two solution concepts, IC and IR constraints are defined accordingly.

\section{A new taxonomy of data pricing methods}
\label{sec:taxonomy}

Data pricing is a challenging problem that has attracted a considerable amount of attention in the literature. To develop a comprehensive understanding of this topic, we now propose a new taxonomy of  existing studies, which is shown in Table~\ref{tab:taxonomy}.


\begin{table*}[h]
  \centering
  \caption{A new taxonomy of data pricing methods. We first divide the studies into three categories, ``Buy side'', ``Sell side'' and ``Two sides''. On Buy side, the studies are further divided by privacy notion, while on Sell side, the studies are divided by query type. Both privacy notion and query type are used to further divide the studies in Two side.We will examine existing studies on data pricing under each of the classes in this classification. The cells in the table indicate the sections in which the respective scenarios are discussed. }
  \label{tab:taxonomy}
  \begin{tabular}{c|c|c|c|c|c}
    \hline
    \multicolumn{3}{c|}{\multirow{3}{*}{\backslashbox{Buy side}{Two sides }{Sell side}}} & \multicolumn{3}{c}{Query type} \\
    \cline{4-6}
    \multicolumn{3}{c|}{~} &
    Null query & One-off query & General query \\
    \cline{4-6}
    \multicolumn{3}{c|}{~} &
    \ref{sec:sell-null} & \ref{sec:sell-general} & -- \\
    \hline
    \multirow{2}{*}{Privacy notion} & Ad hoc & \ref{sec:buy} & \ref{sec:two-null} & -- & -- \\
    \cline{2-6}
     & DP & -- & -- & \ref{sec:two-oneoff} & \ref{sec:two-general} \\
    \hline
\end{tabular}
\end{table*}

\subsection{Market structure}
Our taxonomy is based on market structure. 
Specifically, in a sell-side market, it is the utility of the derived information that determines the prices of data. 
For example, when a retailer makes a sales decision based on people's shopping histories, the monetary value of the shopping history data could be measured in terms of the expected additional profit that is to be generated as a consequence of this decision.
Some metrics are used to measure the goodness of information, such as quality, volume, and accuracy. 
In contrast, in a buy-side market, the value of data lies in their potential monetary loss for data owners due to privacy leakage. 
For example, when someone releases their income data to another person, there is a risk that this data may become accessed by others or even the general public, leading to potential harm to the data owner. The value of data in this case, therefore, should reflect the amount of potential harm (in monetary term) that is expected to incur to the data owner as a result of the release of the data. The data broker often determines the prices of data according to the data owners' privacy preferences and privacy loss they suffer. 
Moreover, in a two-sided market, data reflect the value in the two forms at the same time. A data broker sells data to data consumers for the information value while compensates data owners for their privacy loss. A data broker needs to balance the privacy loss and the information utility, or the cost on data procurement and the profit from data selling. Due to the essential difference of data value on two sides, the studies on data pricing with different market structures should be distinguished. Therefore, we first divide the data pricing studies by market structure.

\subsection{Sell side: Query type}
The classification criterion for the sell side is query type.
For {\em null queries}, data consumers are allowed to access the entire datasets. A data consumer may freely explore the contents of the dataset once the dataset is purchased. In this case, a number of metrics, such as quality and size,  can be used to indicate the value of the dataset. 
For {\em one-off queries}, a data consumer is allowed to issue one fixed query, which returns an aggregate statistic, such as sum, average, minimum, and maximum of the data, as the query answer to the data consumer. The accuracy of the query result directly reflects its value. 
For {\em general queries}, unlike one-off queries where the data consumer only obtains specific query result, the data consumer may issue multiple queries and the query answers can be aggregate statistics or data views consisting of multiple rows and columns. A data consumer can explore the information contained in the view and infer some information by combining multiple query answers.  


\subsection{Buy side: Privacy notion}

The classification criterion on the buy side is a privacy notion rooted in the ability of an individual or a group of individuals to control the access to their personal data.
People have various understandings of privacy and several privacy notions have been proposed.

{\em Differential privacy} is the most extensively adopted privacy notion in the existing studies on data pricing. Basically, differential privacy defines privacy by limiting the disclosure of individual private information from the aggregate information when answering a query on a dataset. 
To such effect, we consider a randomised mechanism $M$, which implements a query on a dataset $D$ and outputs a real-valued aggregate result randomly, i.e., $M(D): \mathbb{R}^n \times \Omega \to \mathbb{R}$, where $\Omega$ is a probability space. Two datasets $D$ and $D'$ that differ on at most one element are called neighbouring datasets. If $\varepsilon$ is a nonnegative parameter, 
$\varepsilon$-differential privacy is defined as follows \cite{Dwork2006Differential}: 
A randomised mechanism $M$ is $\varepsilon$-differentially private if for any pair of neighbouring datasets $D$ and $D'$, and for all possible subsets $S$ in the output space $O$ of $M$:
\[\frac{\Pr \left[ M(D) \in S \right]}{ \Pr \left[ M(D') \in S \right] } \leq e^{\varepsilon}\]
Given two neighbouring dataset $D$ and $D'$, when the ratio above is small, the probability of outputting the same results are close and, thus, the mechanism protects individual privacy. The pamameter $\varepsilon$ can be interpreted as a quantifier of privacy.

In contrast to differential privacy, which has become a standardised privacy notion, several privacy notions are used in an ad hoc way, such as, anonymisation \cite{Xu2015Privacy}, the number of data consumers who are allowed to access the data \cite{gkatzelis2015pricing} and the rate of disclosing the ground truth data \cite{parra2018optimized}. We call these privacy notions as {\em ad hoc notions}. These ad hoc notions define privacy in specific data trading scenarios.
Take the one used in \cite{gkatzelis2015pricing} as an example. A data broker sells a sample dataset to each data consumer. When the data is selected more times and, thus, is accessed by more data consumers, it is more likely the privacy of the corresponding data owner is disclosed. 
This notion, however, can not be adopted to general scenarios.

\bigskip

In the rest of the paper, we first examine work on one-sided markets before moving on to two-sided markets. The reviewed literature is summarised in Table~\ref{tab:summary}. Table~\ref{tab:summary} shows that versioning, pay-per-use and tiered pricing are used as pricing methods in the context of complete information, while auction, contract and incentive mechanism are used in the context of asymmetric information. Further, in the context of complete information, pay-per-use, tiered pricing and versioning are often used for null queries, while versioning is used for general queries.  

\section{Sell-side market}
\label{sec:sell}
The common setup of a sell-side market is as follows: given a data market with a monopolistic data broker and multiple data consumers, the data broker sells queries to the data consumers and aims to achieve certain goals by properly pricing datasets for data consumers. 
Under different query types, data consumers explore information in different ways. We first discuss data pricing studies under null queries and then those under general queries.

\subsection{Sell-side market under null queries}\label{sec:sell-null}


In a sell-side market under null queries, the monopolistic data broker sells null queries to data consumers and aims to achieve goals, e.g., maximising profit, by properly pricing the datasets. Data consumers have heterogeneous preferences towards the datasets, e.g., quantity, quality, valuation or privacy level. In the contexts of complete information, the data broker knows their preferences (see \cite{kushal2012pricing,Yu2017Data}), while in the contexts of asymmetric information, the broker does not know (see \cite{riederer2011sale,mehta2019sell,li2014pricing}).

Kushal et al. \cite{kushal2012pricing} consider a sell-side market with complete information. They investigate the situation where data consumers would like to buy a number of datasets (with homogeneous properties such as size and quality) from a monopolistic data broker and will be charged for each purchase. The goal of the data broker is to set the proper unit price for each dataset based on the WTP of the consumers to maximise profit. 
The authors apply two simple pricing methods, linear and tiered pricing. 
The work assumes that data consumers may experience different marginal utilities for a dataset: One group experiences a constant marginal utility while another experiences a diminishing marginal utility. 
When the cumulative WTP is larger than the total price, the consumer makes the purchase. The total profit is the sum of payments from all data consumers. The optimal prices can be derived from the total profit function. 

In addition to quantity in \cite{kushal2012pricing}, Yu and Zhang \cite{Yu2017Data} also consider the quality of data that affects data prices. The quality of data can be measured by multiple dimensions, such as accuracy, completeness, consistency and timeliness, all contributing to the utility gained by the data consumer \cite{batini2009methodologies}. 
The study considers a scenario where a data broker intends to maximise profit when dealing with a group of utility-maximising data consumers, who have heterogeneous expectations for data quality in each dimension. 
This heterogeneity prompts the data broker to provide multiple versions with different data qualities at different prices. To model the behaviours of the data broker and data consumers, a bi-level programming model involving a monopoly and multiple potential data consumers is proposed. By solving this model, the optimal quality level of each dimension and the price of each version can be derived. The managerial insight of this work is that considering multiple dimensions of datasets better segments the market and increases the profits of data brokers as a consequence. 

Riederer et al. \cite{riederer2011sale} consider a sell-side market with asymmetric information, specifically, the data consumers have hidden information about their data valuation. For the data broker, once a dataset has been produced, it incurs negligible cost to be released to any number of data consumers. The authors thus consider the dataset as having unlimited supply and use an unlimited supply auction based on the {\em exponential mechanism} (see \cite{mcsherry2007mechanism}) for data pricing. The auction guarantees approximate IC. 
%


Mehta et al. \cite{mehta2021sell} consider the case where asymmetric information is about data consumers' demands on datasets. Many data marketplaces provide customised datasets to suit data consumers' need and specifications. That is, each data consumer may specify  an ``ideal'' record, a vector that describes properties of an entry in a desirable dataset. It is assumed that the data consumers can accommodate slight differences between the actual and the ideal datasets but the utility decays as the actual dataset deviates from the ideal one.
Furthermore, both the ideal record and the decay rate are known only by the data owner. The broker who has the entire dataset needs to consider which records to provide and how to set the price to maximise her revenue. The problem is formulated as a multi-dimensional mechanism design problem, which is intractable \cite{daskalakis2014complexity}. The work solves the problem with the assumption that the ideal record is uniformly distributed. The optimal solution is that the data broker gives the records within a certain distance from the ideal record and charges a price that is the difference between the data consumer's utility and the information rent. They also propose an approximate algorithm to solve the problem without this assumption. 

Li and Raghunathan \cite{li2014pricing} also consider the heterogeneous demands of data consumers. Some consumers request ``aggregate'' dataset with low sensitivity while others request ``personal'' dataset with high sensitivity. The desired data types are unknown by the data broker.a situation where the data broker sells two types of datasets, an ``aggregate'' dataset with low sensitivity, and a ``personal'' dataset with high sensitivity.   The pricing of data needs to take into account the data consumers' desired data type, which is hidden to the data broker. To reveal data consumers' desired data type, the data broker offers two forms of contracts, one for each data type. Each contract is a two-part-tariff contract that consists of a fixed part and a variable part. The pricing function is written as $p(s,m)=\alpha_s+\beta_sm$, where $\alpha_s$ and $\beta_s$ are the fixed charge and the variable charge expressed in terms of the sensitivity level $s$ and data quantity $m$. Solving an optimisation problem with IR and IC constraints, one obtains the optimal pricing parameters $\alpha_s$, $\beta_s$ and data quantity $m$, thus solving for the data price. The framework of trading aggregate and personal datasets provides practical insights for data consumers and data brokers. Data consumers can reduce their costs and privacy intrusion by more utilising aggregate data. The data brokers can improve their products by providing two types of datasets accordingly. 


Naghizadeh and Sinha \cite{naghizadeh2019adversarial} consider potential misbehaviours of data consumers. Once a data consumer gets the query answer, there is a chance that the data consumer may attempt to misuse the data (e.g., leaking the privacy of data owners). To prevent this, the authors design contracts for the data trade with multiple types of honest data consumers and a type of adversarial data consumers. Each type of data consumers is given an offer that indicates the price of the issued query and the fine to be paid if the consumer is found misusing the query answer. The data broker designs contracts that cater for different consumers' needs and deter the adversarial data consumers from misusing their data. The problem of contract design for such data trade is formulated as an optimisation problem to maximise the expected revenue subject to IC and IR constraints. The authors then propose an approximation algorithm to solve the problem. 


\subsection{Sell-side market under general queries}\label{sec:sell-general}

In sell-side market under general queries, the common setup is: the monopolistic data broker sells general queries to the data consumers and aims to design pricing functions that satisfy certain properties. When data consumers are allowed to conduct general queries, the data consumer would get, as a query result, a fraction of the dataset that satisfies the query conditions. This fraction is called a {\em view}. 
When a data consumer obtains a view of the dataset, the data consumer may explore further information that is contained in the view. 
There is a chance that the view output by one query may contain the view of another query. For example, the view that contains the incomes of the entire population would contain also the incomes of males, and the incomes of females. Now suppose that the price $p_1$ charged for the incomes of males, and the price $p_2$ charged for the incomes of females add up to a lower value $p_1+p_2$ than the price charged for the incomes of the entire population, the data consumer can purchase the incomes of males and of females, and resell the views of the entire population to gain a profit.  This is an example of an {\em arbitrage}. The notion of ``query-based pricing scheme” is introduced to alleviate this problem \cite{balazinska2011data}.



Formally, given a dataset $D$, explicit prices $p_1,\ldots,p_m$, and  views $v_1,\ldots,v_m$ of $D$, a {\em data query  pricing scheme} $S$ consists of the set of pairs of the form $(v_i,p_i)$ \cite{Balazinska2013A}. 
A view $v$ may be determined by a combination of multiple views from $S$, say $u_1,\ldots,u_\ell$, represented as $u_1,\ldots,u_\ell \twoheadrightarrow v$. A data consumer's  queries can be answered by a view or multiple views and the prices can be automatically derived from the pricing scheme $S$. Such query-based pricing scheme can be seen as versioning in a sense that each view answering a query corresponds to a version of the entire dataset. 

Considering a view $v$ that is determined by a set of views $v_1,\ldots,v_m$, a data consumer buys $v$ at the price $p$ (or buy $v_1,\ldots,v_m$ at price $p_1+\ldots + p_m$). If the sum of prices of $v_1,\ldots,v_m$ was lower than the price of $v$, i.e., $p_1+\ldots+p_m<p$, no one would buy $v$ and all consumers would buy the view set instead and could even resell the view set at price $p$. A key principle in designing pricing functions for queries is {\em arbitrage-freeness} \cite{Balazinska2013A}. Formally, a {\em query-based pricing function} $p(\varphi)$ takes a dataset $D$ and a query $\varphi$ (or a set of queries $\varphi_1,\ldots,\varphi_m$) as inputs and returns a non-negative real number price. Given a query $\varphi$, which is determined by a set of queries, $\varphi_1,\ldots,\varphi_m$, i.e., $\varphi_1,\ldots,\varphi_m \twoheadrightarrow \varphi$, a query-based pricing function is  {\em arbitrage-free} if it satisfies $p(\varphi) \leq \sum_{i=1}^m p_i (\varphi_i)$. 

Most of the studies on data pricing in sell-side market under general queries attempt to address the arbitrage problem by designing arbitrage-free query-based pricing functions. Koutris et al. \cite{koutris2012query} initiate the studies by considering a specific query type, generalised chain queries. Subsequent studies \cite{Lin2014On,Deep2016The,deep2017qirana,deep2017qiranadem,xia2018arbitrage,Tang2013The,upadhyaya2016price,chawla2019revenue} generalise the results to arbitrary queries. 

In addition to arbitrage-freeness, Koutris et al. \cite{koutris2012query} claim that a query-based pricing function should be {\em discount-free}. This means that, with query price $p(\varphi)$, no price of any combinations of queries determining $\varphi$ should be lower than $p(\varphi)$, i.e., when $\varphi_1,\ldots,\varphi_m \twoheadrightarrow \varphi$, $p(\varphi) \geq \sum_{i=1}^m p_i (\varphi_i)$. Otherwise, the data consumers will never buy the combination of queries, $\varphi_1,\ldots,\varphi_m$.
The authors restrict the issued queries to a subclass of conjunctive queries called {\em generalised chain queries}. For example, $\varphi_1 (x,y)= R_1(x),R_2(x,y),R_3(y)$, where $\varphi_1$ is a query with variables $x$ and $y$, and $R_1$, $R_2$ and $R_3$ are tables, is a generalised chain query. Further, they propose a polynomial-time algorithm, which turns generalised chain queries into chain queries and reduces the problem of pricing chain queries into a minimum cut problem. The capacity of an edge is the explicit price of a view and the price of a query can be derived by computing the cost of the minimum cut in the graph. 
This work \cite{koutris2012query} provides insights for data brokers that they should be aware of the new type of arbitrage in data marketplaces and it provides one feasible pricing method to address the arbitrage. 

The pricing methods in \cite{koutris2012query} is designed for a specific query type, generalised chain queries.  
However, in practice, data consumers may issue arbitrary queries such as select-project-join, aggregate and random sample queries, and the ideal pricing functions should be compatible with these queries, which is considered in \cite{Lin2014On}. Lin and Kifer \cite{Lin2014On} suggest that the price of a query reflects how much the query reveals about the authentic dataset and they study the data pricing problem from a probabilistic view. When an answer to a query has a high probability of revealing the entire dataset, it should be set at a high price. Based on this intuition, the authors investigate two pricing schemes for arbitrary queries: {\em answer-dependent pricing} and {\em instance-independent pricing}. In the answer-dependent pricing mechanism, the price of the answer to a query is a function of both queries and answers while in the instance-independent pricing mechanism, the price of a query does not depend on the database instance.
Let $\mathcal{D}$ be the set of all database instances. For any  $D\in \mathcal{D}$ and answer $\omega \in range(\varphi)$, the probabilities of the answer to a query $\varphi$ being $\omega$ is $\Pr [\varphi(D)=\omega]$.
The answer-dependent pricing function is $p(\varphi,\omega)=\sum_{D\in \mathcal{D}} w_D \left(1- \frac{\Pr \left(\varphi(D) =\omega \right)}{\max_{D' \in \mathcal{D}} \Pr \left[\varphi(D')=\omega \right]} \right)$, where $w_D$ is a non-negative weight for each database instance $D$.
The instance-independent pricing function is $p(\varphi)=\sum_{D \in \mathcal{D}} \sum_{\omega \in range(\varphi))} w_D \Pr[\varphi(D)=\omega]  \log \frac{\Pr \left[\varphi(D)=\omega \right]}{\sum_{D'\in \mathcal{D}} \Pr \left[\varphi (D')=\omega \right]}$, where $w_D$ is a non-negative weight such that $\sum_{D\in \mathcal{D}} w_D=1$. These two pricing functions are proved to be arbitrage-free. 

In the same vein, Deep and Koutris \cite{Deep2016The} 
give a complete characterisation of arbitrage-free, instance-independent pricing functions and arbitrage-free, answer-dependent pricing functions.
In an answer-dependent pricing scheme, if a data consumer issues a query $\varphi$ with answer $\omega$, she would realise that the dataset $D$ cannot be any $D' \in \D$ where $\varphi(D') \neq \varphi(D)$. The collection of all such database instances $\overline{\mathcal{S}}_{\varphi}(\omega)=\{D' \in \D\mid \varphi(D')\neq \omega\}$ is called a {\em conflict set}, which determines the price of the query. Given a database instance $D$, set $\mathcal{S}_D=\{\overline{\mathcal{S}}_{\varphi}(\varphi(D))\mid \varphi \in \Phi\}$, where $\Phi$ is the set of queries, which forms a join-semilattice under the partial ordering $\subseteq$. 
A pricing function $p(\varphi,\omega)=f(\overline{\mathcal{S}}_{\varphi}(\omega))$ is arbitrage-free if and only if $f$ is monotone and subadditive over every semilattice $(\mathcal{S}_D,\subseteq)$ where $D\in \D$.
Examples of such pricing functions include the weighted coverage function $\sum_{D:\varphi(D)\neq \omega}w_D$, and the supremum function $\sup _{D:\varphi(D)\neq \omega}w_D$, where $w_D$ is the weight assigned to each $D\in \D$. 
Similarly, in instance-independent pricing, 
for a query $\varphi \in \Phi$, the authors define a partition $\mathcal{P}_{\varphi}$ on $\D$ induced by the equivalence relation: $D \sim D'$ if and only if $\varphi(D)=\varphi(D')$. If every block in a partition $\mathcal{P}_{\varphi_1}$ is a subset of some block in another partition $\mathcal{P}_{\varphi_2}$, then $\mathcal{P}_{\varphi_1}$ refines $\mathcal{P}_{\varphi_2}$.
The partitions ordered by the refinement relation form a join-semilattice.   
The pricing function $p(\varphi)=f(\mathcal{P}_{\varphi})$ is arbitrage-free if and only if $f$ is monotone and subadditive over this join-semilattice.
Examples of such pricing functions include Shannon entropy function and q-entropy function. 

Computing the price of a query using the pricing functions in \cite{Deep2016The} over the set of all possible datasets is a $\# P$-hard problem.
To address this computational limitation, Deep et al. \cite{deep2017qirana,deep2017qiranadem} propose a framework named QIRANA, which applies the pricing functions in \cite{Deep2016The} but looks at only a smaller set $\mathcal{C} \subseteq \D$. This set $\mathcal{C}$ is called a support set and every member of $\mathcal{C}$ needs to have its price predefined by the data broker. In QIRANA, $\mathcal{C}$ consists of a certain number of randomly chosen database instances that differ from the true instance $D$ at one or two rows. Using this set, QIRANA can efficiently compute the price of a query over a large-scale dataset. Moreover, QIRANA can also support {\em history-aware pricing} \cite{upadhyaya2016price}. A database is inherently dynamic as new tuples are inserted into the database through time. A data consumer, on the other hand, may make a number of queries on the same database at different time instances. Even though these queries may be the same, the answers may be different due to the dynamic nature of the database. However, these queries have certain overlap and it is possible that the consumer pays for the unchanged part of the data repeatedly. History-awareness requires that a data consumer is charged only once for any view she has purchased. QIRANA in \cite{deep2017qiranadem} treats the new query and all queries in the history as a query bundle and use the arbitrage-free pricing functions on the query bundle to set the price. 



The previous studies on query-based data pricing address the potential arbitrage problem by proposing well-designed pricing functions. However, none of them consider the revenue optimisation aspect of query-based pricing schemes. Chawla et al. \cite{chawla2019revenue} take the revenue into consideration. They consider data trade between a data broker and multiple data consumers, each of them wants to call a query and has a budget for it. Only when the price of the query is within the budget, the consumer will buy it. Further, since the reproduction cost of data is negligible, the data broker has an unlimited supply. The problem for the data broker is to maximise its revenue, which is the sum of the payments from all data consumers who can afford the queries, subject to the pricing function being monotone and subadditive. This problem is formulated as an optimisation problem.


Xia and Muthukrishnan \cite{xia2018arbitrage} defines a new type of arbitrage, called version-arbitrage. In some cases, two queries $\varphi_1,\varphi_2$ produce very similar views of the database. For example, when querying about the income of all people above $20$ years old, and that of all people above $21$ years old, the returned views can be very close. If the price of the former is higher than that of the latter, a data consumer who intends to issue the first query would buy the second one instead since the tuples in the second view are definitely in the first view. 
Formally, for two queries $\varphi_i,\varphi_{i'}$, denote the ratio between the number of tuples satisfying both $\varphi_i$ and $\varphi_{i'}$ and those satisfying $\varphi_{i'}$ by $\Pr(\varphi_i|\varphi_{i'})$. A query-based pricing function is version-arbitrage-free if $p(\varphi_{i'})\geq \Pr(\varphi_i|\varphi_{i'}) p(\varphi_{i})$. The authors propose query-based pricing functions that guarantee version-arbitrage-freeness.
One pricing function is the uniform pricing function that assigns each query a uniform price. The other pricing function is a non-uniform pricing function that assigns varying prices for different queries. They also present a greedy algorithm to find out the optimal prices. It starts with an optimal uniform pricing and iteratively updates the price for one query while preserving the version-arbitrage-freeness property. The main insight of this work is that it proposes a new definition of arbitrage and provides a more flexible way for data broker to sell queries. 

In some applications, pricing an entire view is seen as too coarse and it makes sense to price individual tuples in a dataset. Tang et al. \cite{Tang2013The} tackle this problem and propose a query pricing model based on tuples. They apply the {\it data provenance} techniques to find the minimal set of tuples which contribute to a query answer, which is called minimal provenance. Data provenance is the description of the origins of a piece of data and its processing history \cite{Buneman2004Why}. Given a database $D$ and a query $\varphi(D)$, the provenance is denoted by $L$  ($L \subseteq D$). In the pricing model of \cite{Tang2013The}, given a set of tuples $X=\{t_1,\ldots,t_k\} \subseteq D$, the price of $X$ is defined by the $p$-norm as $||X||_p=\left(\sum_{i=1}^m s_D (t_i)^p \right)^{\frac{1}{p}}$, where $t_i\in X$ and $s_D$ is a price setting function. Given a query $\varphi$ on a database $D$, the set of tuples used to produce the query answer $\varphi(D)$ is defined as the set of  minimal provenances $L_{\min} (\varphi,D)$. Then the price of a query $\varphi(D)$ is determined by its cheapest minimal provenance, $p_D (\varphi)=\min_{L \in L_{\min} } ||L||_p$.

\section{Buy-side market}
\label{sec:buy}
The common setup of a buy-side market is as follows: given a data market with a monopolistic data broker and multiple data owners, the data broker procures private data from data owners, and aims to achieve certain goals by properly compensating data owners. 
The main issue concerns the compensation to data owners' perceived loss of privacy. Privacy notion matters in terms of the way of quantifying privacy. The reviewed papers studying buy-side market use ad hoc privacy notion, and none of them adopts differential privacy. Specifically, privacy is measured by the probability of disclosing sincere data in \cite{parra2018optimized}, by the privacy protection level realised by anonymisation in \cite{Xu2015Privacy}, and by the number of data consumers who are allowed to access data in \cite{gkatzelis2015pricing}. 

%

Data owners have different privacy attitudes or preferences. Such information is known by the data broker in the context of complete information (See \cite{parra2018optimized}), and hidden in the context of asymmetric information (See \cite{Xu2015Privacy,gkatzelis2015pricing}).   

Under complete information assumption, Parra-Arnau \cite{parra2018optimized} treats private data as information products and applies versioning as the pricing method. Different from classical information products, data have an attribute of privacy. The data owners have private data in multiple categories, such as health, religion, among others and attaches different privacy preferences to $m$ different categories. The privacy level for each category $j\in \{1,\ldots,m\}$ is captured by parameter $\delta_j$, which indicates the probability of disclosing the authentic data. Further, the data owner declares her compensation requirement $p_j$ for revealing the authentic data of each category and the total payment is $p=\sum_{j=1}^{m} \delta_j p_j$. The question for the data owner is, given the payment $p$, how to design an optimal version that minimises her privacy disclosure by setting proper values of $\delta_j$ where $j\in\{1,\ldots,m\}$. The authors formulate the problem as an optimisation problem that minimises the privacy disclosure and applies Karush-Kuhn-Tucker (KKT) conditions to find out the optimal $\delta_j$, $j\in\{1,\ldots,m\}$, and the price accordingly. The managerial insights of this work include: It provides a way for data owners to fully control their privacy disclosure. It also provides the optimal disclosure strategy, i.e., disclosing the same amount of privacy in each category. 

Under incomplete information, Gkatzelis et al. \cite{parra2018optimized} also allow the data owners to control their privacy disclosure.  
To control privacy disclosure, data owners can set a limit on the number of data consumers who have access to their data. The privacy loss is a function of the number limit. 
In addition, the data owners have different risk attitudes. Taking advantage of different risk attitudes, the authors design an auction mechanism, called certainty equivalent (CE) mechanism, which asks each data owner to report her value and offers two payment options: the first one is a fixed payment, which is a function of her privacy valuation and the number of consumers accessing the data; and the second one is a random payment that depends on the probability that her data is utilised.

Also under incomplete information, Xu et al. \cite{Xu2015Privacy} argue that the value of data is determined by the utilities to data consumers, and the utilities are negatively related to privacy protection, as a more stringent privacy implies a greater distortion or omission in the purchased data. Given a data owner’s privacy preference, the data broker provides a menu of contracts including three elements, which represent the privacy level, the utility that the data consumer gains, and the transfer to the data owner, respectively. All of these three elements are determined by the data owner's privacy preference. To identify the optimal contract, the authors model the problem as a maximisation problem with the goal of maximising the data broker’s expected utility under IC and IR constraints and a data broker’s minimum requirement on the total utility she gains from data. The optimal transfer function and production function are derived from the optimisation model and the value of privacy level is determined by experiments. 
This work provides managerial insights for data brokers on designing contract for data trading. Specifically, forming the optimal contract largely depends on the utility required by data consumers and the utility of data to data consumers.

\section{Two-sided market}
\label{sec:two}
The common setup of a two-sided market is as follows: given a monopolistic data broker, multiple data owners, and multiple data consumers, the data broker procures private data from data owners and sells queries to data consumers. 
The data broker essentially faces two data-pricing tasks, one on data owners and the other on data consumers. Thus, the data broker needs to balance the considerations of data privacy of the data owner and utility of the data consumer simultaneously. This is the biggest difference between a two-sided market and a one-sided market, and poses the major challenge for this scenario. When dealing with the two sides at the same time, both the privacy notion on the buy side and the query type on the sell side are considered. According to these two criteria, the existing studies on two-sided markets are classified into two-sided markets under null queries and ad hoc privacy notion, two-sided markets under one-off queries and differential privacy, and two-sided market under general queries and differential privacy. 

\subsection{Two-sided market under null queries and ad hoc privacy notion}\label{sec:two-null}


In a two-sided market under null queries and ad hoc privacy, the data broker sells null queries to data consumers and procures private data from data owners using ad hoc privacy notions to quantify privacy. 
The four works in this category study a two-sided market with incomplete information. To be specific, Jaisingh et al. \cite{jaisingh2008privacy} and Zheng et al. \cite{Zheng2017Trading} assume that the data owners' valuation is unknown. Cao et al. \cite{Cao2017An} assume that valuations of data owners and data consumers are unknown. Bergemann et al. \cite{bergemann2022economics} assume that the private data of data owners is unknown. 


Zheng et al. \cite{Zheng2017Trading} consider a two-sided market where the data broker knows the data valuations of data consumers, but does not know those of data owners. Specifically, they consider a crowd-sensed data market where there are multiple data acquisition points at a certain place and multiple data owners at each point. The data broker aims to 
maximise the revenue from data consumers while minimising the cost on data collection at the same time. 
The solution to this problem consists of two parts, one for revenue maximisation while the other for cost minimisation. To maximise the revenue, it is assumed that the expected payment for each data acquisition point is known, and a greedy algorithm is proposed to select the most cost-effective acquisition point. To minimise the cost, an auction is used to select data owners at each point. 

Cao et al. \cite{Cao2017An} consider a two-sided market where the data valuations of all participants are unknown, and address the problem from the perspective of a system designer who aims to maximise the social welfare by properly allocating data resources among data owners, data broker and data consumers. Here social welfare is defined as the difference between the data consumers' utility, and the sum of the dis-utility to data owners due to privacy loss, and the cost of data collection to the data broker. 
The authors propose an iterative auction to solve this problem. In an iterative auction, each time a new auction round begins, an auctioneer announces the data allocation, pricing and compensation rules with the goal of maximising the social welfare. Then every bidder, be it a data owner, a data consumer, or a data broker, calculates a bid that can maximise her utility, and then submits the bid with the required amount of data. Using the submitted bids, the auctioneer then sets the rules for the next round of the auction. This task is formulated as a constrained optimisation problem. Once the rules are derived, a new round begins. This auction continues until social welfare converges.
The auction mechanism is proved to be individually rational and weakly budget feasible. 

Jaisingh et al. \cite{jaisingh2008privacy} consider a different two-sided market such that a data market is embedded in a digital platform where the service provider collects the data from its users for third party data consumers. In other words, the service provider is the data broker, while the users are the data owners. On one hand, the data broker needs to make decisions on data collection and the price of the service; on the other hand, she needs to determine the price for data consumers. The authors apply contracts to the problem where an Internet service provider (ISP) acts as a data broker. They divide the Internet users into two types according to their privacy cost, being high or low. In relation with the users, the data broker has three data collection and pricing strategies, (1) provide a single contract that specifies the price of Internet service under a data collection policy; (2) provide a single Internet service contract where data are not collected; (3) offer both contracts above. The optimal strategy is determined by the relationships among privacy loss, the privacy cost of two types of users and the total gain from the data trade, which provides insights to ISPs.

Bergemann et al. \cite{bergemann2022economics} also consider such a two-sided data market. They take the correlation between the data of data owners into account. That is, some information of non-users of the service can be inferred from the data collected from the users and thus the non-users also suffer from privacy loss. The authors propose a bilateral contract between the users and the data broker, and between the third party and the data broker. 
The contract defines a data sharing policy,  either to share complete data (with identity information) or to share aggregate data (without anonymised information), and corresponding prices. 
The optimal contract with a producer is to share the complete data and charge the increased profit due to the data, while the optimal contract with a user is to collect anonymised data and compensate her with the difference between the utility for others and the utility for herself, which provides managerial insights for data brokers on designing contracts. A similar problem setup is also discussed in \cite{ichihashi2021economics,acemoglu2022too}.

\subsection{Two-sided market under one-off queries and differential privacy}
\label{sec:two-oneoff}

In a two-sided market under one-off queries and differential privacy, the monopolistic data broker sells a single query answer that a data consumer is interested in and procures private data from data owners using differential privacy to quantify privacy. 
According to the predefined differential privacy, 
some noise has to be added to the query answer, which in turn compromises the accuracy of the query answer and the information utility to data consumers. Therefore, when setting procurement prices of data for data owners, a data broker should consider the trade-off between the privacy protection for data owners and the information utility of data consumers. The data owners with higher privacy requirements should be compensated less and the data consumers with higher accuracy requirements should be charged more. Further, on the buy side, data owners have heterogeneous privacy preferences, and, thus, are willing to sell data at different prices. 
However, the valuation is hidden but useful information for data brokers. All of the studies in this category are under the incomplete information assumption.  
The studies in this category are initiated by a seminal study of Ghosh and Roth \cite{ghosh2011selling} and extended in \cite{fleischer2012approximately,ligett2012take,nissim2014redrawing,roth2012conducting,dandekar2012privacy,zhang2020selling,cummings2015accuracy,ghosh2014buying,wang2016buying} by considering different aspects of one-off queries.

Ghosh and Roth \cite{ghosh2011selling} view privacy as the unit to be priced (i.e., in a $\varepsilon$-differentially private mechanism, $\varepsilon$ units of privacy are traded) and apply a multi-unit procurement auction. Specifically, they consider the problem where a data broker would like to answer a count query by collecting data from multiple data owners, each of whom is associated with a private data entry $d$ and a valuation $\theta$ for privacy. The privacy cost $c$ of a data owner with valuation $\theta$ when the data is used under $\varepsilon$-differential privacy is assumed to be a linear function, i.e., $c=\varepsilon \theta$. Better privacy protection often means less accuracy of the query results and, thus, the data broker needs to balance the trade-off between privacy protection and query accuracy. The authors consider how to set procurement prices in two settings, (1) maximisation of the accuracy subject to data broker's budget constraint, and (2) minimisation of the total payments with a fixed accuracy goal. In the first setting, the FairQuery mechanism is proposed, which first sorts privacy valuation in an ascending order, and then selects as many data owners as possible within the budget, paying a uniform monetary compensation to each of the selected data owners. 
In the second setting, the number of winners is determined by the accuracy requirement. The MinCostAuction mechanism sorts the costs of data owners in ascending order. Here, the cost is the product of the privacy valuation and the privacy parameter. The compensation for each of the selected data owner equals the cost of the first unselected one. To protect privacy, noises drawn from Laplace distributions are added to the two mechanisms, respectively. The mechanisms guarantees dominant strategy IC, IR, ex-post BF and accuracy. The managerial insight of this work is that to achieve a certain privacy level, data brokers need to purchase from at least a certain number of data owners with privacy higher than a certain level. 

Ghosh and Roth \cite{ghosh2011selling} show a negative result when there is a unknown correlation between the private data $d$ and privacy valuation $\theta$, no individually rational direct mechanism can protect privacy and achieve non-trivial query accuracy. An example of the correlation is that a patient with diabetes tends to attach higher valuation to her health data. 
The impossibility result is reconsidered and bypassed by using indirect mechanisms \cite{roth2012conducting,fleischer2012approximately,ligett2012take} or setting an extra assumption \cite{nissim2014redrawing}. 

Roth and Schoenebeck \cite{roth2012conducting} allow arbitrary correlation between the private data and the privacy valuation and reconsider the problem of \cite{ghosh2011selling} in a Bayesian setting where the valuation distribution is a known prior. They propose a take-it-or-leave-it offer with a lottery mechanism which is defined by a distribution $\mathcal{G}$. For each data owner $i$, a random price $p_i$ is drawn from $\mathcal{G}$. When the reported valuation $\theta_i$ is smaller than $p_i$, the probability of using the data entry is $1$ and the payment is a function of the variance level and the reported valuation; otherwise, the probability of buying the data is zero and so is the payment. Due to the randomness of the data owner selection, no one can infer any private information from whether a data entry is utilised even under the assumption of the correction between the data entry value and the privacy valuation. One issue that comes from the lottery mechanism is that the query answer might be biased since the selecting probability of data owners with different valuations varies. To address such issue, the authors of \cite{roth2012conducting} apply Horvitz-Thompson estimator \cite{horvitz1952generalization}, which scales up the data value by the inverse selecting probability and, thus, ensures the estimator is unbiased.   
The mechanism in \cite{roth2012conducting} guarantees Bayesian IC, ex-post IR, interim BF and accuracy. 

In the same vein, Fleischer and Lyu \cite{fleischer2012approximately} reconsider the impossibility results and focus on count queries. To avoid the problem that data consumers may infer some private information just by individual participation decisions based on the correlation between the privacy valuations and the data value, the probability of individuals accepting the offer should be the same in spite of their privacy valuations. They assume that distributions of privacy costs incurred by individuals with different data value are different, and the cumulative distribution function of privacy valuation of individuals with data value $j$ is publicly known and represented by $F_j$. The mechanism firstly find a value $\alpha_j$ so that $F_j (\alpha_j )=pr$, where $pr$ is a predefined threshold, and set the expected payment as $\varepsilon \alpha_j$, where $\varepsilon$ is the privacy parameter, to each individual having data $j$. At last, the mechanism conducts calculations based on the data collected and adds a Laplace noise to the calculation results. The mechanism ensures Bayesian IC, interim IR and accuracy.

Similarly, Ligett and Roth \cite{ligett2012take} design a contract to bypass the impossibility results. Different from \cite{roth2012conducting,fleischer2012approximately}, they do not require Bayesian assumption with the prior knowledge about the privacy valuation distribution. The contract they propose is composed of three elements: the price of the privacy, the privacy protection level for the participation decision and the privacy protection level for the private data. Given a desirable accuracy level, a targeted sample size is set. The offer is provided to each data owner until the number of data owners accepting the offer reaches the target size. Then a Laplace noise is added to the count query answer derived from the sample dataset. 

In contrast to \cite{roth2012conducting,fleischer2012approximately,ligett2012take} who circumvent the impossibility results by using indirect mechanism, Nissim et al. \cite{nissim2014redrawing} propose a direct mechanism but incorporate an extra assumption about the monotonicity of privacy valuation. To be specific, they assume that data owners who have a certain value of private data often attach higher valuation for their privacy. For instance, those who have certain disease tends to be more reluctant to reveal their medical information and, thus, need to be paid more to release their data. They propose a mechanism that works as follows. Given $n$ data owners, a budget $B$ and a privacy protection level $\varepsilon$, for a data owner with a valuation lower than $B/2\varepsilon n$, the mechanism keeps their data unchanged and pays $B/n$ to each data owner; for data owners with high valuations, the mechanism changes their data to $0$ and pays them nothing. Lastly, noise drawn from the symmetric geometric distribution is added. It achieves ex-post IR for all data owners and dominant strategy IR for data owners with low privacy valuations, BF and accuracy.   

The above studies, including \cite{ghosh2011selling,fleischer2012approximately,ligett2012take,nissim2014redrawing}, investigate the pricing problem on a specific query type: count query. However, 
commonly used queries in real world is not confined to count queries. Thus, another extension of \cite{ghosh2011selling} is to consider more general query types \cite{dandekar2012privacy,zhang2020selling}. 

Dandekar et al. \cite{dandekar2012privacy} focus on linear predictor queries.  Linear predictor is to predict the behaviour of a new user in recommender systems by means of exploring the similarity between an existing data owner and a new user. Specifically, a dataset $D$ that consists of personal data $d_i$ of data owner $i$ is assessed with a linear predictor $\varphi(D)=\sum_{i=1}^n w_i d_i$, where $w_i$ is a public weight associated with $i$ to describe the similarity. Data owners have heterogeneous privacy protection requirements and privacy valuations. More weights can make the prediction more accurate, and, thus, more data owners are needed. However, it means a higher payment. Then the pricing problem is formulated as a maximisation problem with the objective of maximising the sum of weights and constraints of budget and accuracy. The problem can be viewed as a 0/1 knapsack problem and a greedy approach is used to solve it. The main insight of this work is parallel with the privacy-accuracy trade off of \cite{ghosh2011selling}. That is, for linear predictor queries, to get a certain level of accuracy, the weights of the data owners with good privacy should be at most a certain proportion of total weight. 

In addition to linear predictor, Zhang et al. \cite{zhang2020selling} allow other common query types, such as count and median queries. Further, the authors reconsider the assumption in the previous studies that data owners are willing to provide their data at any privacy protection level determined by the data broker, and assume single-mind data owners who sell their data only when their privacy requirements are met. The data owners report both their privacy valuations and their privacy requirements to the data broker. Then the pricing problem is formulated as an optimisation problem that maximises query accuracy with the IC, IR and budget constraints and is solved by applying the KKT conditions.  

In \cite{ghosh2011selling}, the data owners may lie about their privacy valuations but can not misreport their data entries since it is assumed that data entries are verifiable. However, in some cases, there is no ground truth dataset and the data broker is not able to verify the data after data collection. Such a scenario is considered in \cite{ghosh2014buying,cummings2015accuracy,wang2016buying}. 

Ghosh et al. \cite{ghosh2014buying} consider the data trade with unverifiable data and design a peer prediction mechanism, which is original for truthfully eliciting a group of experts' opinions on the occurrence rate of some future events. The unverifiable data are thought of as the opinions in a peer prediction mechanism and the data broker would like to predict whether the event would happen according to the data owners' opinions. The authors design a payment rule such that a data owner with low cost would participate the mechanism and maximise her expected utility when her reported data aligns with the result derived from the rest data owners. Thus, the mechanism guarantees truthful report on private data. The main insight of this work is that data brokers can utilise peer prediction mechanisms to address data unverifiability issues. 

Even though the data are not verifiable, the mechanism in \cite{ghosh2014buying} incentivises data owners to report their data truthfully. Cummings et al. \cite{cummings2015accuracy} and  Wang et al. \cite{wang2016buying} consider a slightly different scenario where the data broker is not trustworthy for data owners and data owners are allowed to report noisy data. 

Wang et al. \cite{wang2016buying} apply a local model of differential privacy, where individuals can add noise to their data and exchange the noisy data for monetary compensation. They consider a specific scenario where a data consumer who is interested in a state and data owners have some knowledge, a binary signal, which can indicate the state. The goal of the data consumer is to gain accurate information at a minimal expected payment. They design a payment mechanism where all individuals are asked to report their private data. Then, if one’s data match the majority of others, she will get a reward; otherwise, she will get a penalty. For example, if a data consumer reports $1$ and the state derived from the data of the others’ is $0$, then this data consumer has to pay a penalty. In this mechanism, an equilibrium is that all agents report right signals with a fixed probability, and, thus, it satisfies locally differential privacy.  

Cummings et al. \cite{cummings2015accuracy} allow data owners to give data with noise. A data broker provides data owners multiple options with different variance levels and asks them to report their privacy valuation at each level and aims to minimise the total payment by selecting proper data owners with proper variance level. 
The problem of designing the required mechanism is formulated as a combinatorial optimisation problem whose objective is to minimise the total cost, with constraints on the accuracy. The problem is then transformed into an integer linear program, and further relaxed to a linear program. By solving this linear program, the optimal solution can be found, which is the optimal allocation rule indicating the probability that the data owners are assigned to particular variance levels. Then the Vickrey-Clarke-Groves (VCG)  payment rule \cite{vickrey1961counterspeculation,clarke1971multipart,groves1973incentives} is applied. 

In contrast to the above work that consider the characteristics of datasets, Zhang et al. \cite{zhang2016privacy} consider the positive network effect on the data owners' participation in a data trade. That is, when there is more participants in a data market, the data owners are more willing to get on board and, thus, the privacy cost decreases with the number of participants. For a data owner $i$ with valuation $v_i$, when her data is protected under $\varepsilon_i$-differential privacy, the cost function is modelled as $c_i=\varepsilon_i v_i k^{-t}$, where $k$ is the number of participants and $t\geq 0$ is a parameter used to describe the positive network effect. 
They design an auction which first randomly partitions the data owners into two groups. Then it sorts bids in each group in the increasing order and computes the optimal profit for each group. The lower profit is set as the target profit. For the data owners in the group with higher profit, the first $k$ data owners are selected until the target profit is reached. 

\subsection{Two-sided market under general queries and differential privacy}
\label{sec:two-general}

In a two-sided market under general queries and differential privacy, the monopolistic data broker sells general queries to data consumer and procures private data from data owners using differential privacy to quantity privacy. In addition to the trade-off between privacy protection and payment discussed in the previous subsection, potential arbitrage conditions complicate the data pricing problem even further. On the buy side, the data broker needs to protect the privacy of data owners and properly compensate them for the privacy loss; on the sell side, the data broker needs set proper prices that reflect the information value so that arbitrage is avoided. Such problem is considered in \cite{Li2014A,Nget2017How,niu2018unlocking,niu2019erato,niu2019making}. 

Li et al. \cite{Li2014A} consider a two-sided market where data consumers are allowed to issue multiple linear queries and specify the accuracy requirement. Here the accuracy requirement is represented by a variance level. 
When a query is conducted multiple times, the variance reduces and privacy protection degrades. In the setting with multiple queries under privacy protection, arbitrage-freeness is redefined. It requires that the price of a query with high variance should be cheaper than that of a query with low variance. The authors devise pricing functions and compensation functions respectively for the two sides of a data market. For a data owner, given a variance level and a query specified by a data consumer, the compensation is proportional to the sensitivity of the query function on the change of this data owner's data and inversely proportional to the variance level. The compensation function is proved to be arbitrage free. Then for the data consumer, the price of the query is the sum of the compensations for all data owners involved. The pricing function for a query can be adjusted as long as it is a sub-additive, non-increasing function, such as linear combination, maximum, etc. The pricing functions in such forms are proved to be arbitrage-free. The main insights of this work are it proposes a redefined concept of arbitrage-freeness in the context of privacy-preserving data marketplace and provides useful pricing functions for data brokers. 

Different from \cite{Li2014A}, where data consumers specify the accuracy they expect and the data broker calculates the minimum price, Nget et al. \cite{Nget2017How} consider the data market where data consumers have a budget and the broker returns the most accurate answer.  The data owners have different privacy preferences and are divided into two types: conservative and liberal. To favour these two types of data owners, they propose two compensation functions. The proposed mechanism consists of three parts: randomly choosing a sample of data owners, computing compensations, and selecting the most accurate sample. To be specific, the mechanism firstly chooses a set of data owners with a certain size, which is calculated using a statistical formula. In the second part, given a budget of a data consumer, the money is equally allocated to selected data owners and the privacy levels of all selected data owners are determined using the inverse function of the compensation function. According to the privacy levels, noisy are added to the query result and the accuracy is measured. The sampling process is repeated for multiple times, and the mechanism compares the query results derived from the samples and choose the most accurate one. 

Niu et al. \cite{niu2018unlocking,niu2019erato} consider a two-sided data market where individual data entries are potentially correlated. In other words, the revelation of one data entry may lead to the leakage of other data owners’ privacy. Therefore, even though the data entry of a certain data owner is not utilised, she should be compensated as long as her data entry is correlated with the selected ones. The compensation function should be arbitrage-free and dependently fair. Here dependent fairness requires that when a data owner and her correlated data owners are not involved, the compensation for her should be zero. The authors propose a bottom-up design and a top-down design which both determine the prices for data consumers and the compensations for data owners. The bottom-up design firstly calculates the privacy compensation and then determines the price data consumers should pay while the top-down design works inversely. In the bottom-up design, a pricing scheme similar to that in \cite{Li2014A} is applied. The difference lies in the compensation function. Due to the correlations between data entries, they utilise dependent differential privacy instead of differential privacy to quantify the privacy loss on dependent datasets. 
The compensation is proportional to the dependent sensitivity. Here, dependent sensitivity is defined as the change of the function value due to modification of one data entry and its correlated data entries. 
In the top-bottom design, given the budget of a data consumer, the compensation for each data owner who contributes to the query answer is the proportion of her privacy loss over the total privacy loss of all involved data owners. 
The study is further extended in \cite{niu2019making} by considering the temporal correlations of time-series data, such as continuous monitoring of heart rates and physical activities, when trading time-series data. Niu et al. \cite{niu2019making} deploy the bottom-up design and apply the concept of Pufferfish privacy \cite{kifer2012rigorous} to quantify the privacy under temporal correlations.


\begin{table*}
\centering
\begin{tabular}{lllll}
\hline
References & Market Structure & Privacy notion & Query type & Pricing method \\
\hline
    \cite{kushal2012pricing} & Sell-side & -- & Null query & pay-per-use, tiered pricing  \\   
    \cite{Yu2017Data} & Sell-side & -- & Null query & Versioning  \\
    \cite{riederer2011sale} & Sell-side & -- & Null query & Auction  \\
    \cite{mehta2021sell} & Sell-side & -- & Null query & Incentive mechanism  \\
    \cite{li2014pricing} & Sell-side & --& Null query & Contract\\
    \cite{naghizadeh2019adversarial} & Sell-side & -- & Null query & Contract \\
    \cite{koutris2012query} & Sell-side & -- & General query & Versioning  \\
    \cite{Lin2014On} & Sell-side & -- & General query & Versioning \\
    \cite{Deep2016The} & Sell-side & -- & General query & Versioning  \\
    \cite{deep2017qirana,deep2017qiranadem} & Sell-side & -- & General query & Versioning  \\
    \cite{upadhyaya2016price} & Sell-side & -- & General query & Versioning \\
    \cite{chawla2019revenue} & Sell-side & --  & General query & Versioning \\
    \cite{xia2018arbitrage} & Sell-side & -- & General query & Versioning \\
    \cite{Tang2013The} & Sell-side & -- & General query & Versioning  \\
    \cite{parra2018optimized} & Buy-sided & Ad hoc & -- & Versioning  \\
    \cite{gkatzelis2015pricing} & Buy-side & Ad hoc & -- & Auction  \\
    \cite{Xu2015Privacy} & Buy-side & Ad hoc & -- & Contract \\
    \cite{Zheng2017Trading} & Centralised two-sided & Ad hoc & Null query & Auction  \\
    \cite{Cao2017An} & Centralised two-sided & Ad hoc & Null query & Auction  \\
    \cite{jaisingh2008privacy} & Centralised two-sided & Ad hoc & Null query & Contract \\
    \cite{bergemann2022economics} & Centralised two-sided & Ad hoc & Null query & Contract \\
    \cite{ghosh2011selling} & Centralised two-sided & DP & One-off query & Auction  \\
    \cite{roth2012conducting} & Centralised two-sided & DP & One-off query & Incentive mechanism  \\
    \cite{fleischer2012approximately} & Centralised two-sided & DP & One-off query & Incentive mechanism \\
    \cite{ligett2012take} & Centralised two-sided & DP & One-off query & Contract  \\
    \cite{nissim2014redrawing} & Centralised two-sided & DP & One-off query & Auction   \\
    \cite{dandekar2012privacy} & Centralised two-sided & DP & One-off query   & Auction  \\
    \cite{zhang2020selling} & Centralised two-sided & DP & One-off query & Auction \\
    \cite{ghosh2014buying} & Centralised two-sided & DP & One-off query & Incentive mechanism  \\
    \cite{wang2016buying} & Centralised two-sided & DP & One-off query & Incentive mechanism \\
    \cite{cummings2015accuracy} & Centralised two-sided & DP & One-off query & Incentive mechanism  \\
    \cite{zhang2016privacy} & Centralised two-sided & DP & One-off query & Auction \\
    \cite{Li2014A} & Centralised two-sided & DP & General query & Versioning  \\
    \cite{Nget2017How} & Centralised two-sided & DP & General query & Incentive mechanism  \\
    \cite{niu2019making} & Centralised two-sided & DP & General query & Versioning  \\
    \cite{niu2018unlocking,niu2019erato} & Centralised two-sided & DP & General query & Versioning  \\
\hline
\end{tabular}
\caption{Summary of reviewed literature}
\label{tab:summary}
\end{table*}


\section{Further discussions in economics of privacy}
\label{sec:eco}



The economic value of privacy has been widely investigated in the field of economics both theoretically and empirically. The focus of the studies in economics is to show the value the private data by analysing the consequences of privacy disclosure or privacy protection in varying scenarios and offer managerial insights on private data collection and pricing.   
 
{\bf The value of private data in decision making. }
Posner \cite{posner1981economics} argues that protecting private data may incur inefficiency in marketplaces. That is, the protection of private data makes some potentially useful information unavailable, which may negatively affect the decision making of individuals or organisations who need the information, and then leads to inefficiency in marketplaces as a consequence. Stigler \cite{stigler1980introduction} has the similar argument. As people tend to preserve the privacy of their negative information, the decisions made on the biased information would lead to inefficiency. 

{\bf The value of private data for consumers. }
Varian \cite{varian1996economic} shows that 
consumers, who are also data owners, would like producers to know certain types of their private data such as their preferences, as the consumers could reduce their search costs, while they would like other certain types of private data such as their willing to pay to be secrete. Ali et al. \cite{ali2020voluntary,fainmesser2022digital} show that private data collection is beneficial for consumers.  

{\bf The value of private data for online users. }
Friedman et al. \cite{friedman2001social} investigate the value of privacy in online environment where the users can preserve their privacy by changing identifiers constantly. In such an environment, it is hard for the anonymous users to be trusted and to build cooperation with others, which causes social cost. 
Bapna et al. \cite{bapna2016one} consider online dating websites where the users can anonymously check others' profiles. The empirical results show that the anonymous users have fewer matches than those non-anonymous. 
Gradwohl \cite{gradwohl2017information} considers the effect of privacy protection on social network users. The social network users on one hand enjoy the network effects from participation in the network, on the other hand suffer from privacy loss due to the interdependence among them. The author shows privacy protection increases social welfare in such social networks. 

{\bf The value of private data in price discrimination.} Consumers' electronic footprint such as purchase histories is useful for dynamic pricing, a type of price discrimination. 
Rossi et al. \cite{rossi1996value} show that the utilisation of the purchase histories significantly increases the company's revenue. Similar conclusion is reached by \cite{chen2009dynamic,acquisti2005conditioning,cummings2016strange}. Conitzer et al. \cite{conitzer2012hide} show that using purchase history for price discrimination increases social welfare.

{\bf The value of private data in contract.}  Calzolari and Pavan \cite{calzolari2006optimality} consider an environment where two principles sequentially contract with an agent and show that the share of personal data can enhance social welfare of the three parties. 
Liang and Marsden \cite{liang2020data} 
consider the contract between one principle and an agent where two types of private data are to be utilised, and show that the disclosure of one type increases social welfare while that of the other type reduce welfare. 


{\bf The value of private data in advertising.}
De Corniere and De Nijs \cite{de2016online} consider the value of consumers' private data in sponsored search auctions and show that by disclosing consumers' data to advertisers, advertising platforms increase the quality of match between advertising and consumer demand. 
Goldfarb and Tucker
\cite{goldfarb2011privacy} conduct survey and the results show that the implementation of privacy regulations significantly decreases advertising effectiveness. 

The above studies show the value of private data by analysing the effect of privacy change. Some other studies do this by explicitly quantifying the monetary value of private data. For instance, Hann et al. \cite{hann2007overcoming} show that the data owners in the US value the privacy protection between $\$30.49$ and $\$44.62$. The value of privacy protection is also investigated in  \cite{rose2005data,tsai2011effect,olejnik2013selling}. Huberman et al. \cite{huberman2005valuating} show that the average monetary value of weight and age data are $\$57.56$ and $\$74.06$, respectively. Subsequently, \cite{danezis2005much,cvrcek2006study,carrascal2013your,staiano2014money,zhang2022experimental} investigate the monetary value of location data, browsing histories, mobile data and other types of private data. 


\section{Future opportunities}
\label{sec:future}
Apart from the existing data pricing methods, there are still challenges and opportunities as discussed below:

    (1) The data pricing problem has attracted the attention from many different fields, including but not limited to economics, databases and data management, management science and computer science. The problem of data pricing is often complicated especially when considering the presence of asymmetric information in data marketplaces, which makes the traditional analysis methods invalid or inefficient. Also, data trading is not a one-shot business and it requires dynamics pricing, which further complicates the data pricing problem. Therefore, machine learning and reinforcement learning are powerful assistant tools for addressing those issues. \cite{chen2018dynamic} and \cite{xu2016dynamic} use reinforcement learning to find out the optimal flat fee and the optimal contract, respectively. Machine learning and reinforcement learning have the potential to be applied to other pricing methods.
    
    (2) Most of the existing studies on data pricing focus on the trading of raw datasets and query answers. The essential value of data lies in providing insights to make better decisions and data analytics using machine learning has become ubiquitous in science, business intelligence and many other domains. The trading items should not be restricted to raw datasets or simple queries. Instead, the trading item can be learning models trained on rich data. Chen et al. \cite{chen2019towards} propose a {\em model-based pricing framework} for such a data trade. 
    However, it is investigated by only a few studies, e.g., \cite{zheng2017online,zheng2019arete,chen2019towards,agarwal2019marketplace,abernethy2015low}. A key issue of model-based pricing is how to assess the quality of learning models and the valuation to data consumers, which are essential to set proper prices of learning model and it is open to future research.  
    
    (3) Data can be divided into three types, structured data, unstructured data and semi-structured data. Structure data is often represented in relational tables or statistical data while unstructured data is a sequence of symbols and often coded in natural language. And semi-structure data have structure with some degree of flexibility, such as data in XML \cite{batini2009methodologies}. Most of the studies on data pricing focus on the structured data and rarely discuss the cases for unstructured data, such as data in the format of natural language, social network data and transportation network data. Nevertheless, the latter is the essential input for tasks such as natural language processing and graph representation learning and, thus, it calls for a data trading framework that accommodates unstructured and semi-structured data. 
    
    (4) As Privacy Enhancing Technologies (PETs) evolve rapidly, their adoption and deployment become more widespread. The current studies on data pricing apply the notions of differential privacy and $k$-anonymity. A lot of other PETs deployed in practice, such as homomorphic encryption, multi-party computation, federated analysis, and pseudonymisation, lack investigations in this field. These PETs achieve privacy protection and define privacy in various ways and, thus, require specific attention when designing data pricing frameworks under different privacy definitions. 
    
    (5) The WTP (or WTA) and demand (or supply) of data consumers (or data owners) are not easy to be modelled. For data consumers, data is experience good. Only after purchasing data and deriving useful information, the data consumers can estimate the value of the purchased data. In other words, it is hard for data consumers to predicate the value of data and express their WTP when making purchasing decisions. For data owners, they rarely quantify the value of private data as privacy is intangible. Thus, it is also hard to express their WTA when making selling decisions. Studies such as \cite{huberman2005valuating,hann2007overcoming} empirically investigate the monetary value of private data, which provides insights for modelling WTA and supply of data owners. Investigations on the value of data from the perspective of data consumers are needed.  
    
    (6) Most of the existing studies focus only on the cost for compensating privacy loss of data owners, and ignore other costs, such as the cost for data integration, storage and maintenance, which is not trivial especially for big size of data. The non-trivial costs may affect data collection from data owners and data pricing for data consumers, thus, should be taken into consideration in future work. 

\section{Conclusion}
\label{sec:conclusion}
The increasing demand for data has given rise to the emergence of a novel business model, data marketplaces, and raised a question for both researchers and entrepreneurs: How to set prices of data properly? To address this problem, many scholars have proposed data pricing methods from different perspectives. This paper provides a taxonomy of data pricing methods and a systematic review of existing literature on data pricing. 

We first present the basics of data marketplaces. Then we overview fundamental pricing methods. 
After that, we propose a new taxonomy of data pricing studies with an emphasis on the essential value of data. Data reflects their value in different senses on the two sides of a data marketplace, which makes the market structure the basis of the taxonomy. Then for different query types and privacy notions, data value are measured in different ways. Therefore, we use query type and privacy notion as the criteria on sell side and on buy side, respectively. According to the taxonomy, we provides detailed reviews of existing studies on data pricing in each category and illustrates the applicability and limitations of pricing methods in each category. At last, we outline challenges and opportunities in the field of data pricing. 

To sum up, this survey has meticulously detailed the current status of the research on data pricing methods. It is helpful for us to understand pricing theories in data marketplaces and provides managerial guidance on how to use pricing theories in practice. The survey may become instrumental in pointing at bridges that help close the gap between the approaches to data pricing proposed by academia and the industry practices that have adopted their current pricing methods. In practice, data marketplace pricing predominantly relies on applying simple pricing methods, such as flat-fee, premium and linear pricing. Almost invariably, once a unit data price is determined the amount of data is the only factor that forms the price for a data consumer. Other factors, extensively discussed in this survey such as privacy, security and consumer’s utility, are rarely considered except in the cases where data companies trade with personal data.

\bibliographystyle{ieeetr}
\bibliography{main.bib}

%

\newpage

\begin{IEEEbiography}[{\includegraphics[width=1in,height=1.25in,clip,keepaspectratio]{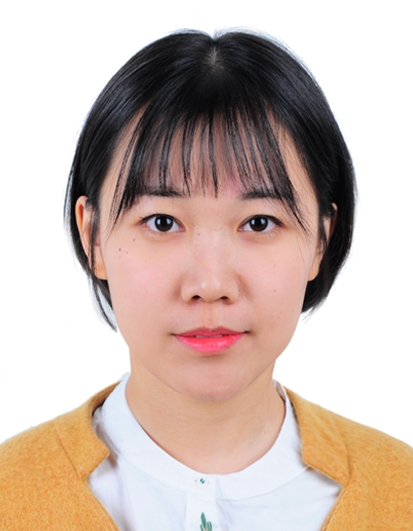}}]{Mengxiao Zhang} (mengxiao.zhang@uestc.edu.cn) is a Research Associate of the Algorithms and Logic research lab at the University of Electronic Science and Technology of China. Her research interests include algorithmic mechanism design and economics of security and privacy.
\end{IEEEbiography}

\begin{IEEEbiography}[{\includegraphics[width=1in,height=1.25in,clip,keepaspectratio]{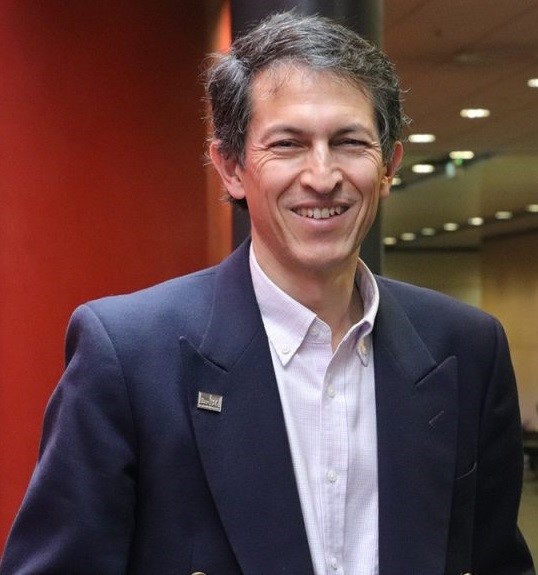}}]{Fernando Beltr\'an} (f.beltran@auckland.ac.nz) is Associate Professor of the Department of Information Systems and Operations Management at the University of Auckland in New Zealand. Fernando received an Electrical Engineering Bachelor's degree from Universidad de Los Andes in Colombia, and Ph.D. in Applied Mathematics from State University of New York in Stony Brook, NY. His current research is on the economics of data privacy protection and data marketplaces, strategic models of competition and cooperation in radio spectrum sharing, and immersive analytics supported by network data visualization. Dr. BELTRAN currently leads the Aroaro Research Space, the UoA data visualization lab.
\end{IEEEbiography}

\begin{IEEEbiography}[{\includegraphics[width=1in,height=1.25in,clip,keepaspectratio]{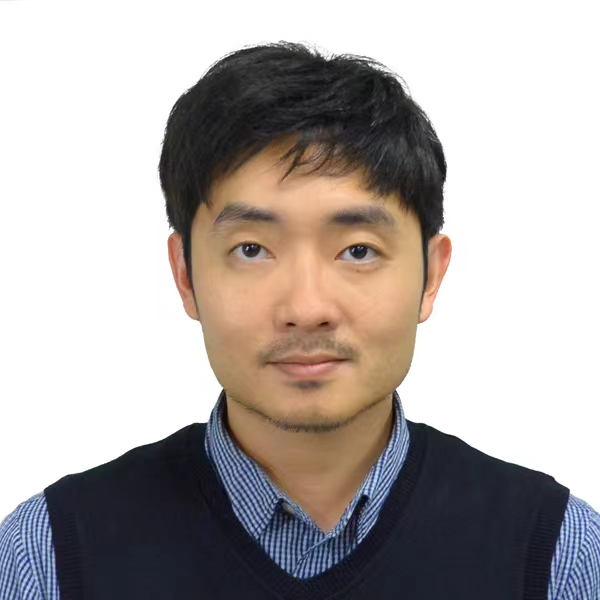}}]{Jiamou Liu} (jiamou.liu@auckland.ac.nz) is a Senior Lecturer in the School of Computer Science at The University of Auckland, New Zealand. He received both his BSc and PhD in Computer Science from The University of Auckland in 2010, and went on to serve as an Associated Researcher at the University of Leipzig in Germany and a Senior Lecturer at AUT, New Zealand. Dr. Liu's research interests lie at the intersection of artificial intelligence, multi-agent systems, and computational social networks. 

\end{IEEEbiography}







\end{document}